\documentclass[aps,prd,eqsecnum,preprint,tightenlines,nofootinbib,showpacs]{revtex4-1}
\usepackage{graphicx,amsmath,latexsym}

\def\bea{\begin{eqnarray}}
\def\eea{\end{eqnarray}}
\def\be{\begin{equation}}
\def\ee{\end{equation}}
\newcommand{\ub}[1]{\underline{#1}}

\begin{document}

\title{Nonperturbative Pauli--Villars regularization \\
of vacuum polarization in light-front QED}

\author{Sophia S. Chabysheva}
\author{John R. Hiller}
\affiliation{Department of Physics \\
University of Minnesota-Duluth \\
Duluth, Minnesota 55812}

\date{\today}

\begin{abstract}

We continue the development of a nonperturbative light-front
Hamiltonian method for the solution of quantum field theories
by considering the one-photon eigenstate of Lorentz-gauge QED.
The photon state is computed nonperturbatively for a Fock
basis with a bare photon state and electron-positron pair
states.  The calculation is regulated by the inclusion
of Pauli--Villars (PV) fermions, with one flavor to make
the integrals finite and a second flavor to guarantee a
zero mass for the physical photon eigenstate.  We compute
in detail the constraints on the PV coupling strengths
that this zero mass implies.  As part of this analysis,
we provide the complete Lorentz-gauge light-front QED
Hamiltonian with two PV fermion flavors and two PV photon
flavors, which will be useful for future work.  The need
for two PV photons was established previously; the need
for two PV fermions is established here.

\end{abstract}

%
\pacs{12.38.Lg, 11.15.Tk, 11.10.Gh, 11.10.Ef
}

\maketitle

\section{Introduction}
\label{sec:Introduction}

The nonperturbative solution of quantum field theories is a very
difficult problem.  For weakly coupled theories, this is usually
avoided, and perturbation theory is applied.  For strongly coupled
theories, in particular quantum chromodynamics, the nonperturbative
problem cannot be avoided for long.  Various nonperturbative
methods have been developed, including lattice theory~\cite{Lattice,TransLattice},
Dyson--Schwinger equations~\cite{DSE}, and light-front Hamiltonian
approaches~\cite{DLCQreview,Glazek,Karmanov,Vary,TwoPhotonQED}, 
and have met with some success.
The light-front methods have the distinct advantage of providing wave
functions as part of the solution.  The wave functions appear as
coefficients in a Fock-state expansion for the Hamiltonian
eigenstate.

Here we continue development of a particular light-front Hamiltonian
method~\cite{bhm,YukawaOneBoson,
OnePhotonQED,ChiralLimit,Thesis,SecDep,TwoPhotonQED}
based on Pauli--Villars (PV) regularization~\cite{PauliVillars}.
Much of the recent development has been in QED~\cite{OnePhotonQED,
ChiralLimit,Thesis,SecDep,TwoPhotonQED}, where results can be checked
against perturbation theory, but which shares the gauge-theory
nature of QCD.  However, there is no expectation of being
able to compete with perturbative QED for accuracy; any but
the lowest-order Fock-space truncations require numerical techniques,
where the accuracy is typically on the order of 1\%.  Thus, the method
is not likely to compete with perturbation theory for any
weakly coupled theory, but this is not a flaw in a method intended
for strongly coupled theories.

The previous work considered eigenstates of
a fermion dressed by one or more scalar or vector bosons.
Eventually we wish to extend the dressed-fermion calculations to 
include one or more fermion-antifermion pairs.  As a first 
step in this direction, we consider the vacuum-polarization
correction to the one-photon state of light-front QED.
The Fock basis is then simply the bare photon state and
the electron-positron states, plus their PV counterparts.
This will allow us to understand how such states can be
included in the dressing of an additional fermion.

The PV regularization method relies upon the introduction
of heavy PV fields to the Lagrangian.  Some are assigned
a negative norm, and the interaction terms are built from
zero-norm combinations of the fundamental fields.  The
negative norms provide the cancellations needed to regulate
perturbation theory, and we find that the nonperturbative
eigenvalue problem is then also regulated.  The use of
zero-norm combinations in the interactions 
eliminates~\cite{YukawaOneBoson} the instantaneous fermion 
contributions~\cite{DLCQreview} from the light-front
Hamiltonian, and, in the case of a gauge theory, allows
the use of gauges other than light-cone gauge~\cite{OnePhotonQED}.
We discuss these features in more detail in the next section.

To regulate the dressed-electron problem, we used one
PV Fermi field and two PV photon fields~\cite{ChiralLimit}.
One of each is sufficient to make the integral equations
finite, but the second PV photon flavor is needed to
maintain the chiral symmetry of the massless-electron limit.
For the present calculation of a photon dressed by an 
electron-positron pair, the PV photon flavors are of no 
particular consequence, but we find that we need two PV 
fermion flavors.  One flavor is again enough to have a finite
result, and the second is needed to maintain a zero mass for
the photon.  A zero mass is not otherwise guaranteed, because
the zero-norm fields in the interaction Lagrangian generate 
flavor-changing currents that break gauge invariance~\cite{OnePhotonQED}.

The addition of a second PV fermion flavor to the older
calculations of the dressed-electron state does not create
any new difficulty, because we can simply take the
infinite-mass limit for this flavor and remove it from the
calculation.  However, a new calculation of the 
dressed-electron state that includes electron-positron pairs
will require the second PV fermion flavor.

As higher and higher Fock sectors are included in a calculation,
the number of PV flavors should not change, in general.  An
exception for QED would be any Fock basis that includes the
possibility of light-by-light scattering.  The breaking of
gauge invariance by the flavor-changing currents should
ruin the usual automatic cancellation of divergences for
this process.  Additional PV fields or an explicit counterterm
will be required, but we do not consider this further here.

Although the number of PV flavors need not change, their 
coupling strengths do need to change as more Fock states
are added~\cite{TwoPhotonQED}.  The conditions of chiral
symmetry for massless electrons and zero mass for photons,
which complete the determination of these couplings, 
become complicated nonlinear equations for the
coupling coefficients.  These typically require iterative
techniques for their solution~\cite{TwoPhotonQED}.
At one loop, the conditions can be solved analytically.

The analysis is done in terms of light-cone
quantization~\cite{DLCQreview,Dirac}.  The coordinates
are $x^\pm=x^0\pm x^3$ and $\vec x_\perp=(x^1,x^2)$,
with $x^+$ chosen as the light-cone time coordinate
and the three-vector of space coordinates written
as $\ub{x}=(x^-,\vec x_\perp)$.  The momentum
conjugate to $x^-$ is $p^+$; therefore, the light-cone
three-momentum is $\ub{p}=(p^+,\vec p_\perp)$.  Dot 
products are given by 
$\ub{p}\cdot\ub{x}=\frac12 p^+x^- - \vec p_\perp\cdot\vec x_\perp$.
The light-cone energy is $p^-$, and evolution in light-cone
time is determined by the light-cone Hamiltonian ${\cal P}^-$.
The mass eigenvalue problem, in a frame where the total 
transverse momentum $\vec P_\perp$ is zero, is given by
${\cal P}^-|\ub{P}\rangle=\frac{M^2}{P^+}|\ub{P}\rangle$.

The primary objective is the solution of this eigenvalue
problem in a Fock basis, with the eigenstate $|\ub{P}\rangle$
expanded in terms of the Fock states with wave functions 
as the coefficients.  The eigenvalue problem becomes a
coupled set of integral equations for the wave functions.
Truncation of the basis makes the coupled system finite.
At very low orders of truncation, the system can be
solved analytically; in general, numerical techniques
are required~\cite{TwoPhotonQED}.

The contents of the remainder of the paper are as follows.
In Sec.~\ref{sec:LightFrontQED}, we summarize the formulation
of light-front QED in Lorentz gauge, extended to include
two PV fermion flavors and two PV photon flavors.  We
then construct the photon eigenstate dressed by
fluctuations to an electron-positron pair in Sec.~\ref{sec:DressedPhoton}
and solve the eigenvalue problem to determine the coupling
coefficients.  Section~\ref{sec:Summary} contains a 
discussion of the results.  An appendix describes the
evaluation of a key integral.

\section{Light-front QED in Lorentz gauge}
\label{sec:LightFrontQED}

The Lorentz-gauge QED Lagrangian, regulated by two PV fermion 
flavors and two PV photon flavors, is
\bea \label{eq:Lagrangian}
{\cal L} &= & \sum_{i=0}^2 (-1)^i \left[-\frac14 F_i^{\mu \nu} F_{i,\mu \nu} 
         +\frac12 \mu_i^2 A_i^\mu A_{i\mu} 
         -\frac{1}{2} \left(\partial^\mu A_{i\mu}\right)^2\right] \\
  && + \sum_{i=0}^2 (-1)^i \bar{\psi_i} (i \gamma^\mu \partial_\mu - m_i) \psi_i 
  - e_0 \bar{\psi}\gamma^\mu \psi A_\mu ,  \nonumber
\eea
where
\begin{equation} \label{eq:NullFields}
  \psi =  \sum_{i=0}^2 \sqrt{\beta_i} \psi_i, \;\;
  A_\mu  = \sum_{i=0}^2 \sqrt{\xi_i}A_{i\mu}, \;\;
  F_{i\mu \nu} = \partial_\mu A_{i\nu}-\partial_\nu A_{i\mu} .
\end{equation}
A subscript of $i=0$ indicates a physical field, and $i=1$ or 2 a PV field.
The $i=1$ fields are chosen to have negative norm.  The mass of the
bare photon $\mu_0$ is zero; the mass of the bare electron $m_0$ is
typically close to the physical electron mass $m_e$ for the range
of PV masses usually considered~\cite{TwoPhotonQED}.

The constants $\beta_i$ and $\xi_i$ control the coupling strengths
of the various fields.  These coupling coefficients must satisfy
constraints for the theory to be consistent.  For $e_0$ to be
the bare charge of the bare electron, we require $\beta_0=1$ and
$\xi_0=1$.  The cancellations necessary to regulate perturbation
theory, which must arise in a sum over flavors of each internal 
line, require that $\sum_i (-1)^i\beta_i e_0^2$ be zero for
a fermion line and  $\sum_i (-1)^i\xi_i e_0^2$ zero for
a photon line.  We therefore require
\be \label{eq:2ndconstraint}
\sum_{i=0}^2(-1)^i\beta_i=0, \;\;
\sum_{i=0}^2(-1)^i\xi_i=0 .
\ee
These also guarantee that the combinations $\psi$ and $A_\mu$
in (\ref{eq:NullFields}) have zero norm.  A third pair of
constraints comes from requiring that the photon eigenstate
have zero mass and that the mass of the electron eigenstate
becomes zero when $m_0$ is set to zero.  Since the first
two pairs of constraints imply $\beta_1=1+\beta_2$ and
$\xi_1=1+\xi_2$, this third pair completes the determination
of the coefficients by providing implicit equations for
$\beta_2$ and $\xi_2$.  In Sec.~\ref{sec:DressedPhoton},
we seek $\beta_2$; for discussion of $\xi_2$, see \cite{TwoPhotonQED}.

The fermion fields $\psi_i$ are decomposed into dynamical and
nondynamical parts $\psi_{i\pm}\equiv\Lambda_\pm\psi_i$ by
the complementary projections 
$\Lambda_\pm\equiv\gamma^0\gamma^\pm/2$~\cite{DLCQreview,LepageBrodsky}.
The nondynamical parts satisfy the following constraints ($i=0$,1,2),
obtained from projecting the Dirac equation with $\Lambda_-$:
\bea  \label{eq:Psii-Constraint}
i(-1)^i\partial_-\psi_{i-}+e_0 A_-\sqrt{\beta_i}\psi_- 
  &=&(i\gamma^0\vec\gamma^\perp)\cdot
     \left[(-1)^i\vec\partial_\perp \psi_{i+}
         -ie_0 \sqrt{\beta_i}\vec A_\perp\psi_+\right] 
     \nonumber  \\
   &&  -(-1)^i m_i \gamma^0\psi_{i+} . 
\eea
Ordinarily, light-cone gauge ($A_-=A^+=0$) would be chosen, so that
the constraint for $\psi_{i-}$ can be solved explicitly.  However,
for the construction of the light-front Hamiltonian, we are interested
in only the combination $\psi_-=\sum_i\sqrt{\beta_i}\psi_{i-}$.
The constraint for $\psi_-$ can be obtained from (\ref{eq:Psii-Constraint})
by first multiplying with $(-1)^i\sqrt{\beta_i}$ and then summing
over $i$, which yields
\be
i\partial_-\psi_-
  =(i\gamma^0\vec\gamma^\perp)\cdot
     \vec\partial_\perp \psi_+
      - \gamma^0\sum_i m_i \sqrt{\beta_i}\psi_{i+}.
\ee
The terms containing the photon field cancel because $\sum_i(-1)^i\beta_i=0$.
The nondynamical field $\psi_-$ can then be constructed from a sum
of $\psi_{i-}$ that satisfy the free-fermion constraint.

The mode expansion for the full Fermi field of the $i$th flavor can be
written as
\be
\psi_i=\frac{1}{\sqrt{16\pi^3}}\sum_s\int \frac{d\ub{k}}{\sqrt{k^+}} 
  \left[b_{is}(\ub{k})e^{-i\ub{k}\cdot\ub{x}}u_{is}(\ub{k})
        +d_{i,-s}^\dagger(\ub{k})e^{i\ub{k}\cdot\ub{x}}v_{is}(\ub{k})
        \right].
\ee
The spinors are~\cite{LepageBrodsky}
\bea \label{eq:LightConeSpinors}
u_{is}(\ub{k})&=&\frac{1}{\sqrt{k^+}}
       (k^+ + \vec\alpha_\perp\cdot\vec k_\perp + \beta m_i)\chi_s , \\
v_{is}(\ub{k})&=&\frac{1}{\sqrt{k^+}}
       (k^+ + \vec\alpha_\perp\cdot\vec k_\perp - \beta m_i)\chi_{-s},
\eea
with
\be
\chi_+=\frac{1}{\sqrt{2}}\left(\begin{array}{c} 1 \\ 0 \\ 1 \\ 0 \end{array}\right),\;\;
\chi_-=\frac{1}{\sqrt{2}}\left(\begin{array}{c} 0 \\ 1 \\ 0 \\ -1 \end{array}\right),
\ee
and the nonzero anticommutators are
\bea  \label{eq:bd-anticommutators}
\{b_{is}(\ub{k}),b_{i's'}^\dagger(\ub{k}')\}
   &=&(-1)^i\delta_{ii'}\delta_{ss'}\delta(\ub{k}-\ub{k}'), \\
\{d_{is}(\ub{k}),d_{i's'}^\dagger(\ub{k}')\}
   &=&(-1)^i\delta_{ii'}\delta_{ss'}\delta(\ub{k}-\ub{k}'). \nonumber
\eea

The mode expansion for the $i$th photon flavor is
\be
A_{i\mu}=\frac{1}{\sqrt{16\pi^3}}\int \frac{d\ub{k}}{\sqrt{k^+}}
  \left[a_{i\mu}(\ub{k})e^{-i\ub{k}\cdot\ub{x}}
        +a_{i\mu}^\dagger(\ub{k})e^{i\ub{k}\cdot\ub{x}}\right] ,
\ee
with the commutator
\be  \label{eq:a-commutator}
{[}a_{i\mu}(\ub{k}),a_{i'\nu}^\dagger(\ub{k}')]
   =(-1)^i\delta_{ii'}\epsilon^\mu\delta_{\mu\nu}\delta(\ub{k}-\ub{k}').
\ee
The metric signature $\epsilon^\mu = (-1,1,1,1)$ is chosen for
Gupta--Bleuler quantization~\cite{GuptaBleuler,GaugeCondition}.
Because we do not use light-cone gauge, there is no constraint
on $A_+=A^-$, and, consequently, there will be no instantaneous
photon interaction term~\cite{DLCQreview} in the Hamiltonian.
The gauge condition $\partial^\mu A_{i\mu}=0$ is implemented
as a projection on the Fock states~\cite{GuptaBleuler,GaugeCondition},
as discussed in \cite{ChiralLimit} and the next section.

We can now construct the light-front Hamiltonian ${\cal P}^-$.
The interaction terms are determined by the spinor matrix 
elements
\bea
\bar u_{is'}(p)\gamma^+ u_{js}(q)&=&2\sqrt{p^+q^+}\delta_{s's} , \\
\bar u_{is'}(p)\gamma^- u_{js}(q)&=&\left\{
  \begin{array}{ll} 
     \frac{2}{\sqrt{p^+q^+}}
        [\vec p_\perp\cdot\vec q_\perp\pm i\vec p_\perp\times\vec q_\perp +m_im_j], &
     s'=s=\pm ,\\
     \mp\frac{2}{\sqrt{p^+q^+}}[m_j(p^1\pm i p^2)-m_i(q^1\pm i q^2)], &
     s'=-s=\mp ,
   \end{array} \right. \nonumber \\
\bar u_{is'}(p)\gamma_\perp^l u_{js}(q)&=&\left\{
  \begin{array}{ll} 
     \frac{1}{\sqrt{p^+q^+}}[p^+(q^l\pm i\epsilon^{lk3}q^k)+q^+(p^l\mp i\epsilon^{lk3}p^k)], &
     s'=s=\pm ,\\
     \mp\frac{1}{\sqrt{p^+q^+}}(m_iq^+-m_jp^+)(\delta^{l1}\pm i\delta^{l2}), &
     s'=-s=\mp ,
   \end{array} \right.  \nonumber \\
\bar u_{is'}(\ub{p})\gamma^\mu v_{js}(\ub{q})
   &=&\left(\bar v_{js}(\ub{q})\gamma^\mu u_{is'}(\ub{p})\right)^*
   =\left.\bar u_{is'}(\ub{p})\gamma^\mu u_{js}(\ub{q})
               \right|_{m_j\rightarrow -m_j}^{s\rightarrow -s} , \\
\bar v_{is'}(\ub{p})\gamma^\mu v_{js}(\ub{q})&=&
   \left.\bar u_{is'}(\ub{p})\gamma^\mu u_{js}(\ub{q})
               \right|_{m_j\rightarrow -m_j,\, m_i\rightarrow -m_i}
                          ^{s\rightarrow -s,\, s'\rightarrow -s'} .
\eea
These generalize matrix elements given in \cite{LepageBrodsky} to the case
of unequal masses, to accommodate the flavor-changing currents.  The 
Hamiltonian is then found to be
\bea \label{eq:QEDP-}
{\cal P}^-&=&
   \sum_{i,s}\int d\ub{p}
      \frac{m_i^2+p_\perp^2}{p^+}(-1)^i
          b_{i,s}^\dagger(\ub{p}) b_{i,s}(\ub{p}) \\
   && +\sum_{i,s}\int d\ub{p}
      \frac{m_i^2+p_\perp^2}{p^+}(-1)^i
          d_{i,s}^\dagger(\ub{p}) d_{i,s}(\ub{p}) \nonumber \\
   && +\sum_{l,\mu}\int d\ub{k}
          \frac{\mu_l^2+k_\perp^2}{k^+}(-1)^l\epsilon^\mu
             a_{l\mu}^\dagger(\ub{k}) a_{l\mu}(\ub{k})
          \nonumber \\
   && +\sum_{i,j,l,s,\mu}\sqrt{\beta_i\beta_j\xi_l}\int d\ub{p} d\ub{q}\left\{
      b_{i,s}^\dagger(\ub{p}) \left[ b_{j,s}(\ub{q})
       V^\mu_{ij,2s}(\ub{p},\ub{q})\right.\right.\nonumber \\
      &&\left.\left.\rule{1.75in}{0in}
+ b_{j,-s}(\ub{q})
      U^\mu_{ij,-2s}(\ub{p},\ub{q})\right] 
            a_{l\mu}^\dagger(\ub{q}-\ub{p})  \right. \nonumber \\
&& +     b_{i,s}^\dagger(\ub{p}) \left[ d_{j,s}^\dagger(\ub{q})
       \bar V^\mu_{ij,2s}(\ub{p},\ub{q}) 
+ d_{j,-s}^\dagger(\ub{q})
      \bar U^\mu_{ij,-2s}(\ub{p},\ub{q})\right] 
            a_{l\mu}(\ub{q}+\ub{p})
                      \nonumber \\
&& -   \left.   d_{i,s}^\dagger(\ub{p}) \left[ d_{j,s}(\ub{q})
       \tilde V^\mu_{ij,2s}(\ub{p},\ub{q}) 
+ d_{j,-s}(\ub{q})
      \tilde U^\mu_{ij,-2s}(\ub{p},\ub{q})\right] 
            a_{l\mu}^\dagger(\ub{q}-\ub{p})
                    + H.c.\right\}\,,  \nonumber
\eea
The vertex functions $V$ and $U$ are as given in \cite{OnePhotonQED}:
\bea \label{eq:vertices}
    V^0_{ij\pm}(\ub{p},\ub{q}) &=& \frac{e_0}{\sqrt{16 \pi^3 }}
                   \frac{ \vec{p}_\perp\cdot\vec{q}_\perp
                      \pm i\vec{p}_\perp\times\vec{q}_\perp
                       + m_i m_j + p^+q^+}{p^+q^+\sqrt{q^+-p^+}} , \\
    V^3_{ij\pm}(\ub{p},\ub{q}) &=& \frac{-e_0}{\sqrt{16 \pi^3}}
                        \frac{ \vec{p}_\perp\cdot\vec{q}_\perp
                      \pm i\vec{p}_\perp\times\vec{q}_\perp
                       + m_i m_j - p^+q^+ }{p^+q^+\sqrt{q^+-p^+}} , \nonumber\\
    V^1_{ij\pm}(\ub{p},\ub{q}) &=& \frac{e_0}{\sqrt{16 \pi^3}}
       \frac{ p^+(q^1\pm i q^2)+q^+(p^1\mp ip^2)}{p^+q^+\sqrt{q^+-p^+}} , \nonumber\\
    V^2_{ij\pm}(\ub{p},\ub{q}) &=& \frac{e_0}{\sqrt{16 \pi^3}}
       \frac{ p^+(q^2\mp i q^1)+q^+(p^2\pm ip^1)}{p^+q^+\sqrt{q^+-p^+}} , \nonumber\\
    U^0_{ij\pm}(\ub{p},\ub{q}) &=& \frac{\mp e_0}{\sqrt{16 \pi^3}}
       \frac{m_j(p^1\pm ip^2)-m_i(q^1\pm iq^2)}{p^+q^+\sqrt{q^+-p^+}} , \nonumber\\
    U^3_{ij\pm}(\ub{p},\ub{q}) &=& \frac{\pm e_0}{\sqrt{16 \pi^3}}
       \frac{m_j(p^1\pm ip^2)-m_i(q^1\pm iq^2)}{p^+q^+\sqrt{q^+-p^+}} , \nonumber\\
    U^1_{ij\pm}(\ub{p},\ub{q}) &=& \frac{\pm e_0}{\sqrt{16 \pi^3}}
                            \frac{m_iq^+-m_jp^+ }{p^+q^+\sqrt{q^+-p^+}} , \nonumber\\
    U^2_{ij\pm}(\ub{p},\ub{q}) &=& \frac{i e_0}{\sqrt{16 \pi^3}}
                     \frac{m_iq^+-m_jp^+ }{p^+q^+\sqrt{q^+-p^+}} . \nonumber
\eea
The other four vertex functions are related to these by
\bea
\label{eq:BarVertexFunctions}
\bar V_{ij,2s}^\mu(\ub{p},\ub{q})&=&\sqrt{\frac{q^+-p^+}{q^++p^+}}
    \left.V_{ij,2s}^\mu(\ub{p},\ub{q})\right|_{m_j\rightarrow -m_j}, \\
\bar U_{ij,2s}^\mu(\ub{p},\ub{q})&=&\sqrt{\frac{q^+-p^+}{q^++p^+}}
    \left.U_{ij,2s}^\mu(\ub{p},\ub{q})\right|_{m_j\rightarrow -m_j}, \nonumber \\
\tilde V_{ij,2s}^\mu(\ub{p},\ub{q})&=&\sqrt{\frac{p^+-q^+}{q^+-p^+}}
    \left.V_{ij,2s}^\mu(\ub{q},\ub{p})\right|_{m_j\rightarrow -m_j,\,m_i\rightarrow -m_i}, \\
\tilde U_{ij,2s}^\mu(\ub{p},\ub{q})&=&\sqrt{\frac{p^+-q^+}{q^+-p^+}}
    \left.U_{ij,2s}^\mu(\ub{q},\ub{p})\right|_{m_j\rightarrow -m_j,\,m_i\rightarrow -m_i}.
    \nonumber
\eea

The Hamiltonian does not contain any instantaneous fermion terms~\cite{DLCQreview}.
They cancel between physical and PV contributions because they are independent
of the fermion mass and proportional to $(-1)^i\beta_i$ for the $i$th flavor.
The sum over flavors then yields $\sum_i(-1)^i\beta_i=0$.  This is
independent of the gauge choice and does not even require a gauge theory;
the same cancellation happens in Yukawa theory~\cite{YukawaOneBoson}.  The
absence of instantaneous fermion and instantaneous photon contributions is
important for numerical calculations, where such four-point interactions can
greatly increase  the computational load and matrix storage requirements;
this is partial compensation for the increase in basis size brought by the
PV fields.

\section{Dressed photon eigenstate}
\label{sec:DressedPhoton}

We construct the Fock-state expansion for the photon
eigenstate of the light-front Hamiltonian.  This 
requires some discussion of the projection that
implements the gauge condition~\cite{GaugeCondition,ChiralLimit}.
From the eigenvalue problem we obtain coupled equations
for the Fock-state wave functions.  We are interested in
the leading vacuum-polarization contribution and, therefore,
truncate the Fock basis to include only the bare photon state
and single-fermion-pair states.  The requirement that the
physical photon eigenstate have zero mass then completes
the determination of the fermion coupling coefficients $\beta_i$.

\subsection{Gauge Projection}

The gauge condition $\partial^\mu A_{i\mu}=0$ is implemented
as a projection that eliminates one linear combination of
unphysical polarizations and leaves only a zero-norm contribution
from unphysical polarizations that provides for the residual
gauge freedom~\cite{GaugeCondition,ChiralLimit}.  Let $e_\mu^{(\lambda)}(\ub{k})$
be the polarization vectors, with $\ub{k}$ the photon three-momentum
and $\lambda=0$,1,2,3.  They satisfy the orthogonality properties
\be
e^{(\lambda)\mu} e_\mu^{(\lambda')}=-\epsilon^\lambda \delta_{\lambda\lambda'}
   =g_{\lambda\lambda'}
\ee
and, for the physical polarizations $\lambda=1$ and 2,
\be
k^\mu e_\mu^{(\lambda)}=0\;\; \mbox{and} \;\; n^\mu e_\mu^{(\lambda)}=0,
\ee
with $n$ a timelike four-vector that reduces to $(1,0,0,0)$ in
the frame where $\vec k_\perp=0$.  The annihilation operator for
a particular polarization is given by
\be
a_i^{(\lambda)}(\ub{k})=-\epsilon^\lambda e^{(\lambda)\mu}(\ub{k}) a_{i\mu}(\ub{k})
\ee
and satisfies the commutation relation
\be
[a_i^{(\lambda)}(\ub{k}),a_j^{(\lambda')\dagger}(\ub{k}')]
=(-1)^i \delta_{ij} \epsilon^\lambda \delta_{\lambda\lambda'} \delta(\ub{k}-\ub{k}').
\ee

Because the positive-frequency part of the gauge condition is proportional
to $k^\mu a_{i\mu}=(k\cdot n)(a_i^{(0)}-a_i^{(3)})$, the condition can be
implemented by the projection $(a_i^{(0)}-a_i^{(3)})|\psi\rangle=0$ for
all Fock states $|\psi\rangle$.  This projection can be satisfied by
building Fock states with the physical-polarization operators
$a_i^{(1)\dagger}$ and $a_i^{(2)\dagger}$ and the zero-norm
combination $(a_i^{(0)}-a_i^{(3)})/\sqrt{2}$.  The zero norm
guarantees that the projection condition is satisfied.  It also means
that the unphysical polarizations make no contribution to observables;
they instead represent the residual gauge freedom of the Lorentz gauge.
For the present purpose, we do not need to include the unphysical
polarizations at all.

\subsection{Eigenvalue Problem}

With the truncation to at most one electron-positron pair, the Fock-state
expansion for a photon eigenstate with polarization $\lambda=1$ or 2 and total
three-momentum $\ub{P}$ is
\be
|\psi^{(\lambda)}(\ub{P})\rangle
  =\sum_l z_l^\lambda a_l^{(\lambda)\dagger}(\ub{P})|0\rangle
  +\sum_{ijss'} \int d\ub{k} C_{ijss'}^\lambda(\ub{k}) 
         b_{is}^\dagger(\ub{k}) d_{js'}^\dagger(\ub{P}-\ub{k})|0\rangle .
\ee
Here $z_l^\lambda$ is the bare photon amplitude for the $l$th flavor,
and $C_{ijss'}^\lambda(\ub{k})$ is the two-body wave function for an
electron of flavor $i$, spin $s$, and momentum $\ub{k}$, and a
positron of favor $j$, spin $s'$, and momentum $\ub{P}-\ub{k}$.
We will work in a frame where the total transverse momentum
$\vec P_\perp$ is zero.

This dressed photon state is to be an eigenstate of the light-front
Hamiltonian ${\cal P}^-$ with eigenvalue $M_\lambda^2/P^+$.
Of course, for the physical photon, $M_\lambda$ should be zero.
In terms of the wave functions, the eigenvalue problem becomes the
following coupled set of equations:
\bea
\label{eq:1stCoupledEquation}
\lefteqn{\frac{M_\lambda^2}{P^+}z_l^\lambda=
\frac{\mu_l^2}{P^+}z_l^\lambda} && \\
&& +\sum_{ijss'\mu}\int d\ub{k}(-1)^{i+j}\sqrt{\beta_i\beta_j\xi_l}
      C_{ijss'}^\lambda(\ub{k}) e_\mu^{(\lambda)}(\ub{P})
      [\delta_{s's} \bar V_{ij,2s}^{\mu *}(\ub{k},\ub{P}-\ub{k})
        +\delta_{s',-s} \bar U_{ij,-2s}^{\mu *}(\ub{k},\ub{P}-\ub{k})] , 
        \nonumber \\
\lefteqn{\frac{M_\lambda^2}{P^+}C_{ijss'}^\lambda(\ub{k})=
\left(\frac{m_i^2+k_\perp^2}{k^+}+\frac{m_j^2+k_\perp^2}{P^+-k^+}\right)
  C_{ijss'}^\lambda(\ub{k})} && \\
&& +\sum_{k\mu}z_k^\lambda (-1)^k \sqrt{\beta_i\beta_j\xi_k}
\epsilon^\lambda e_\mu^{(\lambda)}(\ub{P}) 
[\delta_{s's}\bar V_{ij,2s}^\mu(\ub{k},\ub{P}-\ub{k})
 +\delta_{s',-s}\bar U_{ij,-2s}^\mu(\ub{k},\ub{P}-\ub{k})] . \nonumber
\eea
We can then solve explicitly for the two-body wave function,
written here in terms of $x\equiv k^+/P^+$,
\be
C_{ijss'}^\lambda(\ub{k})=\epsilon^\lambda\left(\sum_k (-1)^k \sqrt{\beta_i\beta_j\xi_k}\right)
\sum_\mu \frac{P^+ e_\mu^{(\lambda)}[\delta_{s's}\bar V_{ij,2s}^\mu(\ub{k},\ub{P}-\ub{k})
 +\delta_{s',-s}\bar U_{ij,-2s}^\mu(\ub{k},\ub{P}-\ub{k})]}
 {M_\lambda^2-\frac{m_i^2+k_\perp^2}{x}-\frac{m_j^2+k_\perp^2}{1-x}}.
\ee
Substitution into the first equation, (\ref{eq:1stCoupledEquation}),
and use of the vertex functions (\ref{eq:BarVertexFunctions}), yields
\be \label{eq:Eigenproblem}
M_\lambda^2 z_l^\lambda=\mu_l^2 z_l^\lambda
  + m_e^2\sqrt{\xi_l}\epsilon^\lambda I(M_\lambda^2) \sum_k(-1)^k\sqrt{\xi_k}z_k^\lambda ,
\ee
with $m_e$ the physical mass of the electron and
\be \label{eq:I}
I(M^2)=\frac{e_0^2}{8\pi^3}\sum_{ij}(-1)^{i+j}\frac{\beta_i\beta_j}{m_e^2}
   \int \frac{dx d^2k_\perp}{x(1-x)}
\frac{(1-2x)^2 k_1^2+k_2^2+(m_i(1-x)+m_j x)^2}
{\left[M^2 x(1-x)-(m_i^2+k_\perp^2)(1-x)-(m_j^2+k_\perp^2)x\right]}.
\ee
The form given for $I$ is explicitly for the $\lambda=1$ case; however, for
$\lambda=2$, the first two terms in the numerator are replaced by
$k_1^2+(1-2x)^2 k_2^2$, which is actually equivalent due to the symmetry
of the rest of the integrand with respect to the interchange of $k_1$
and $k_2$.  Therefore, $I$ need not carry a polarization label, and
the eigenmasses $M_1$ and $M_2$ are equal, as one would expect.  Also,
the cancellations provided by the PV fermions are sufficient to render
$I(M^2)$ finite.

\subsection{Analytic Solution}

The remaining equation, (\ref{eq:Eigenproblem}), is a $3\times3$
matrix eigenvalue problem
\be
H\vec z^\lambda = \frac{M^2}{m_e^2}\vec z^\lambda,
\ee
where $\vec z^\lambda=(z_0^\lambda,z_1^\lambda,z_2^\lambda)^T$ and
\be
H=\left(\begin{array}{ccc} 
  \mu_0^2/m_e^2+\xi_0 I(M^2) & -\sqrt{\xi_0 \xi_1} I(M^2) & \sqrt{\xi_0 \xi_2} I(M^2) \\
  \sqrt{\xi_0 \xi_1} I(M^2)    & \mu_1^2/m_e^2-\xi_1 I(M^2) & \sqrt{\xi_1 \xi_2} I(M^2) \\
  \sqrt{\xi_0 \xi_2} I(M^2)    & -\sqrt{\xi_1 \xi_2} I(M^2) & \mu_2^2/m_e^2+\xi_2 I(M^2)
  \end{array} \right) .
\ee
When the bare photon mass $\mu_0$ is zero, the determinant of $H$ is
\be
{\rm det}H=\xi_0\frac{\mu_1^2\mu_2^2}{m_e^4}I(M^2).
\ee
Therefore, the physical photon eigenstate has zero mass, within
the given truncated Fock basis, if and only if $I(0)$ is
zero.  This provides the condition for determination of the
coupling coefficient $\beta_2$.

The integrals in $I(0)$ are simple enough to permit its analytic
evaluation.  This is presented in the Appendix, with the result
that
\be
I(0)=\frac{e_0^2}{8\pi^2} \sum_{ij}(-1)^{i+j}\beta_i\beta_j I_{ij} ,
\ee
with the $I_{ij}$ given in (\ref{eq:Iij}).

\begin{figure}[ht]
\vspace{0.2in}
\centerline{\includegraphics[width=15cm]{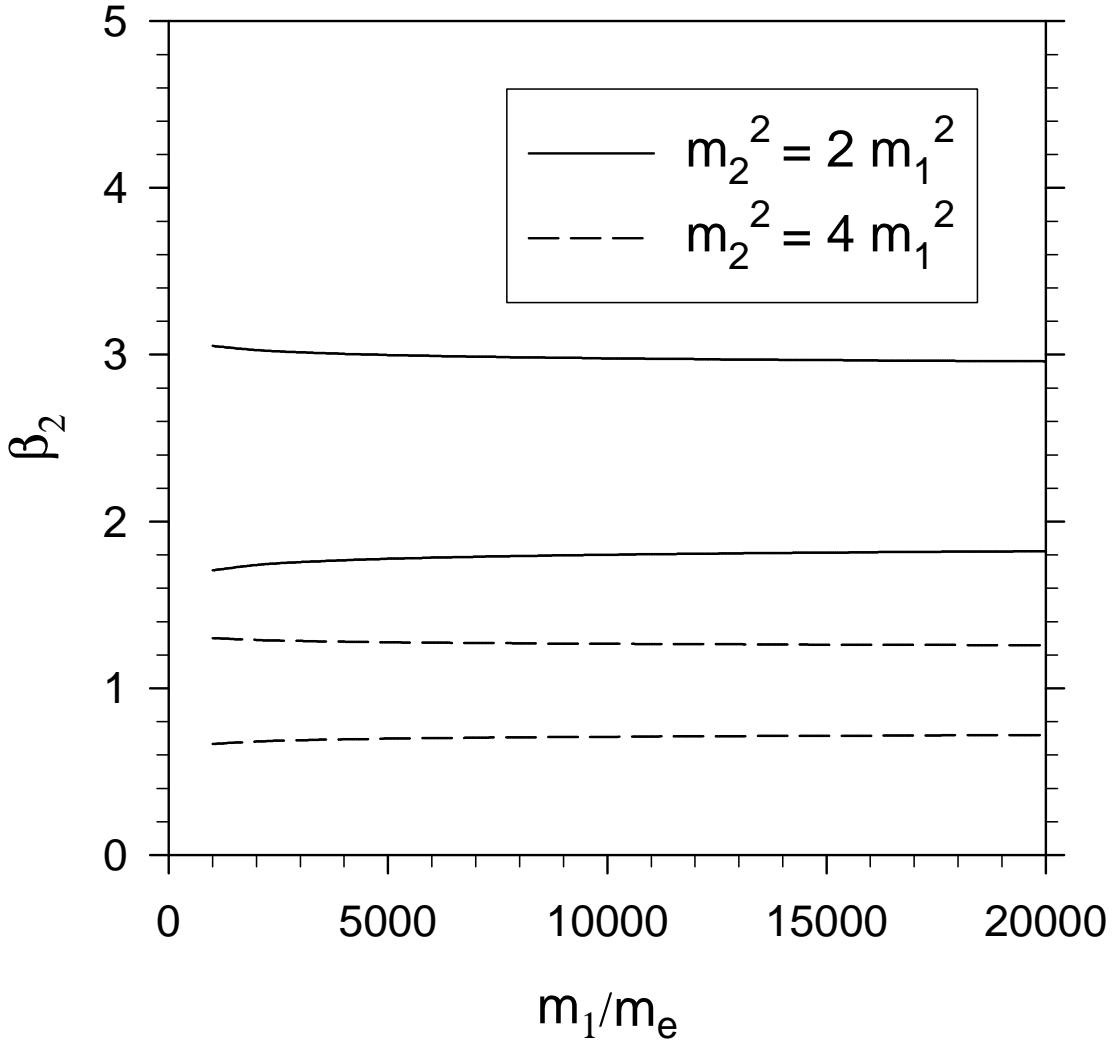}}
\caption{\label{fig:Roots} The coupling coefficient $\beta_2$
as a function of the PV masses $m_1$ and $m_2$.  The two possible
values of $\beta_2$ are determined by the constraint of 
having a zero mass for the physical photon eigenstate.
The value used for the bare electron mass $m_0$ is $0.99 m_e$,
where $m_e$ is the physical electron mass.}
\end{figure}

To use $I(0)=0$ to find $\beta_2$, we replace $\beta_0=1$ and 
$\beta_1=1+\beta_2$, and take advantage of the symmetry $I_{ij}=I_{ji}$,
to write $I(0)=0$ as
\be
I_{00}+I_{11}-2I_{01}+2(I_{11}+I_{02}-I_{01}-I_{12})\beta_2
+(I_{11}+I_{22}-2I_{12})\beta_2^2=0.
\ee
The two roots of this quadratic equation are plotted in Fig.~\ref{fig:Roots}
as functions of the PV masses $m_1$ and $m_2$, with the bare electron
mass set to a typical value for the dressed-electron problem~\cite{TwoPhotonQED}.
The range in $m_1$ is taken up to the point where the earlier
calculations were done for the dressed-electron state~\cite{TwoPhotonQED};
the value of $m_2$ is fixed in ratio to $m_1$.
The eventual choices of the root and 
of the $m_2/m_1$ ratio will be determined by optimization of
the numerical calculation.  Ideally, the root and the ratio
will not be too large; a large root would mean large couplings
for the PV particles, and a large ratio would make $m_2$ yet
another mass scale in the problem.

The main point here is the existence of values of $m_2$ and 
$\beta_2$ for which the mass of the photon eigenstate is zero.
Also, since $\beta_2=0$ is not a root, the addition of the
second PV fermion flavor is necessary to restore the
zero mass.  For calculations in QED that include a single
electron-positron pair in the basis, with no photons in
the same Fock state, the analytic results given here
provide the value to use for $\beta_2$.

\section{Summary}
\label{sec:Summary}

We have shown that the addition of a second PV fermion flavor is
sufficient to restore the physical photon eigenstate to zero
mass.  The photon self-energy induced by vacuum polarization
is thus not only rendered finite by the PV regularization, but
an additional finite correction can also be made by adjusting
the coupling coefficients of the PV fermions.  For the
simplest Fock-state basis, we have computed explicitly the
coupling coefficients as functions of the electron's bare
mass and the PV fermion masses; the results are illustrated
in Fig.~\ref{fig:Roots}.

This analysis provides building blocks necessary for the
extension of previous work on the dressed-electron 
state~\cite{OnePhotonQED,TwoPhotonQED} to include 
electron-positron pairs.  The complete Lorentz-gauge
light-front Hamiltonian (\ref{eq:QEDP-}) has been constructed
and the one-photon eigenstate has been investigated in some
detail.  The issues that remain to be resolved are the 
vacuum-polarization contribution to charge renormalization
and the electron-positron pair contribution to current
covariance.  There are also technical issues to be
addressed, associated with the numerical analysis of
the coupled equations of the dressed-electron eigenproblem.
The size of the calculation will be larger than the case of
two-photon truncation~\cite{TwoPhotonQED}, because
the number of PV fermion flavors will be two instead of one,
but the size should still be small enough for the calculation
to be done.

\acknowledgments
This work was supported in part by the Department of Energy
through Contract No.\ DE-FG02-98ER41087.

\appendix

\section{Evaluation of $I(0)$}  \label{sec:appendix}

We evaluate the integral $I(M^2)$, defined in (\ref{eq:I}),
for the case of $M=0$.
The symmetry of the integrand allows us to replace
$k_1^2$ and $k_2^2$ in the numerator by $k_\perp^2/2$.  The integral
over azimuthal angle can be done immediately, to replace $d^2k_\perp$ by
$\pi dk_\perp^2$.  The numerator can be written as
\be
(1-2x)^2 k_\perp^2/2+k_\perp^2/2+(m_i(1-x)+m_j x)^2
=[m_i^2(1-x)+m_j^2 x+k_\perp^2]-x(1-x)[(m_i-m_j)^2+2k_\perp^2].
\ee
The first bracket can be dropped, since it cancels the matching
bracket in the denominator, leaving an integrand independent of
$i$ and $j$, for which the sums over $i$ and $j$ are zero.  This 
reduces the form of $I(0)$ to
\be
I(0)=\frac{e_0^2}{8\pi^3}\sum_{ij}(-1)^{i+j}\frac{\beta_i\beta_j}{m_e^2}
   \int dx d^2k_\perp \frac{(m_i-m_j)^2+2k_\perp^2}{m_i^2(1-x)+m_j^2 x+k_\perp^2} .
\ee
Further simplification comes from writing
$k_\perp^2=[m_i^2(1-x)+m_j^2 x+k_\perp^2]-[m_i^2(1-x)+m_j^2 x]$
in the numerator and again dropping the first bracket, for
the same reason as before.

The expression that we actually integrate is, then
\be
I(0)=\frac{e_0^2}{8\pi^2} \sum_{ij}(-1)^{i+j}\beta_i\beta_j
\int \frac{dx dk_\perp^2}{m_e^4}
\frac{(m_i-m_j)^2-2[m_i^2(1-x)+m_j^2 x]}
      {(1-x)m_i^2/m_e^2+x m_j^2/m_e^2+k_\perp^2/m_e^2} .
\ee
The $k_\perp^2$ integral yields 
$\ln[(1-x)m_i^2/m_e^2+x m_j^2/m_e^2+k_\perp^2/m_e^2]$
evaluated at 0 and $\infty$; the sums over $i$ and $j$ eliminate
the contributions at the upper limit.  The remaining expression is
\be
I(0)=\frac{e_0^2}{8\pi^2} \sum_{ij}(-1)^{i+j}\beta_i\beta_j I_{ij} ,
\ee
with
\be 
I_{ij}\equiv \int_0^1 \frac{dx}{m_e^2}
  \left\{2[m_i^2(1-x)+m_j^2 x]-(m_i-m_j)^2\right\}
      \ln[(1-x)m_i^2/m_e^2+x m_j^2/m_e^2] .
\ee
When $i=j$, the integrand is trivial; when $i\neq j$, we can use
the transformation $z=(1-x)m_i^2/m_e^2+x m_j^2/m_e^2$ to
arrive at a simple integral.  The final results are
\be \label{eq:Iij}
I_{ij}=\left\{ \begin{array}{ll} 
  2\frac{m_i^2}{m_e^2}\ln\left(\frac{m_i^2}{m_e^2}\right), & i=j \\
  \frac{m_i^2+m_j^2}{2m_e^2}
  -\frac{2m_im_j}{m_e^2}
  +\frac{m_i m_j}{m_j^2-m_i^2}
      \left[\frac{m_i(m_j-2m_i)}{m_e^2}\ln\left(\frac{m_i^2}{m_e^2}\right)
            -\frac{m_j(m_i-2m_j)}{m_e^2}\ln\left(\frac{m_j^2}{m_e^2}\right)\right], & i\neq j .
            \end{array} \right.
\ee
The form of $I(0)$ is now fully specified.


\end{document}